\documentclass[conference, 10pt]{IEEEtran}

\usepackage{cite}
\usepackage{url}
\usepackage{graphicx}
\usepackage{color}
\usepackage{placeins}
\usepackage{float}
\usepackage{tabularx,colortbl}
\usepackage{ifthen}
\usepackage{physics}

\newcommand{\pati}[1]{}

\makeatletter

\newcounter{author}
\renewcommand{\author}[2][]{
   \stepcounter{author}
   \@namedef{author@\theauthor}{#2}
   \@namedef{authorlabel@\theauthor}{#1}
}

\newcounter{address}
\newcommand{\address}[2][]{
   \stepcounter{address}
   \@namedef{address@\theaddress}{#2}
   \@namedef{addresslabel@\theaddress}{#1}
}

\newcommand{\alsep}{and}
\newcommand{\sss}{\vspace{-1.6mm}}

\def\newmaketitle{\par%
  \begingroup%
  \normalfont%
  \def\thefootnote{}
  \def\footnotemark{}
  \let\@makefnmark\relax
  \footnotesize
  \footnotesep 0.7\baselineskip
  \normalsize%
  \twocolumn[\thenewmaketitle\@IEEEaftertitletext]%
  \if@IEEEusingpubid
     \enlargethispage{-\@IEEEpubidpullup}%
  \fi
  \endgroup
  \setcounter{footnote}{0}\let\maketitle\relax\let\@maketitle\relax
  \gdef\@thanks{}%
  \let\thanks\relax}

\def\thenewmaketitle{
  \newpage
  \begin{center}%
    \vskip0.2em{\Huge\@IEEEcompsoconly{\sffamily}\@IEEEcompsocconfonly{\normalfont\normalsize\vskip 2\@IEEEnormalsizeunitybaselineskip
   \bfseries\large}\@title\par}\vskip1.0em\par%
    \vspace{1ex}
    \newcounter{c@author}
    \newcounter{c@tmp}
    \ifthenelse{\value{author}=2}{%
      \newcommand{\liand}{ and }}{%
      \newcommand{\liand}{, and }}
    \ifthenelse{\value{address}<2}{%
      \@nameuse{author@1}%
      \stepcounter{c@author}%
      \whiledo{\value{c@author}<\value{author}}{%
        \setcounter{c@tmp}{\value{author}}%
        \addtocounter{c@tmp}{-\value{c@author}}%
        \ifthenelse{\value{c@tmp}=1}{%
          \renewcommand{\alsep}{\liand}}{\renewcommand{\alsep}{, }}%
        \stepcounter{c@author}\alsep \@nameuse{author@\thec@author}}\\%
    }
    {
      \@nameuse{author@1}${}^{(\ref{\@nameuse{authorlabel@1}})}$%
      \stepcounter{c@author}%
      \whiledo{\value{c@author}<\value{author}}{%
      \setcounter{c@tmp}{\value{author}}%
      \addtocounter{c@tmp}{-\value{c@author}}%
      \ifthenelse{\value{c@tmp}=1}{%
        \renewcommand{\alsep}{\liand}}{\renewcommand{\alsep}{, }}%
      \stepcounter{c@author}\alsep \@nameuse{author@\thec@author}%
        ${}^{(\ref{\@nameuse{authorlabel@\thec@author}})}$%
      }
    }
    \vspace{0.2ex}

    \ifthenelse{\value{address}>0}{%
      \ifthenelse{\value{address}=1}{
        {\@nameuse{address@1}}
      }
      {
        \newcounter{c@address}

        \begin{center}
        \whiledo{\value{c@address}<\value{address}}
        {
          \refstepcounter{c@address}
            ${}^{(\thec@address)}$\,%
              \label{\@nameuse{addresslabel@\thec@address}}%
              \@nameuse{address@\thec@address}\\ %
        }
        \end{center}
      } 
    }
    {
      \relax
    }
  \end{center}
}

\makeatother

\title{Electron Scattering at a Soft \\ Temporal Potential Step}

\author[org1]{Furkan Ok}
\author[org1]{Christophe Caloz}

\address[org1]{Department of Electrical Engineering, KU Leuven, Kasteelpark Arenberg 10, 3001, Leuven, Belgium (furkan.ok@kuleuven.be)}

\begin{document}

\newmaketitle


\begin{abstract}
We solve the problem of electron scattering at a soft temporal potential step. Given the relativistic nature of the problem, we use the Dirac equation, with its spinor wavefunction. We find solutions in terms of hypergeometric functions, which demonstrate that the observed phenomenon of later forward-wave and backward-wave electron scattering previously reported for a sharp (Heaviside) temporal potential step can be obtained experimentally and applied to new electronic devices.
\end{abstract}

\section{Introduction}\sss
\pati{GSTEMs}
Generalized Space-Time Engineered-Modulation (GSTEM) metamaterials, or GSTEMs for short, represent one of the latest developments in metamaterial science and technology~\cite{Caloz_2023_GSTEM}. They are made of metaparticles formed by modulation interfaces between media of different parameters that move as perturbations without any net motion of matter. As such, they still produce typical moving-media effects, such as Doppler shifting, Fresnel-Fizeau dragging, and frequency chirping, without requiring cumbersome moving parts.

\pati{Emergence of Q-GSTEMs}
To date, reported GSTEMs have been almost exclusively restricted to \emph{classical} structures, or C-GSTEMs. A few notable exceptions include~\cite{schultheiss_2021_time,Dong_2024_Qtime}, which investigated embryonic forms of \emph{Quantum} GSTEMs, or Q-GSTEMs. Given the huge diversity and immense potential of that nascent field, it seems quite safe to predict at this point a brilliant future to Q-GSTEMs, both in terms of scientific and technological innovations. The problem of electron scattering at a sharp temporal potential step, recently reported in~\cite{Ok_arXiv_2023}, is an example of a canonical GSTEM problem. However, a sharp step discontinuity is non-causal!

\pati{Contribution}
In this paper, we consider a \emph{causal}, and hence realizable, form of the electron scattering problem introduced in~\cite{Ok_arXiv_2023}: we transform the sharp (Heaviside) potential step into a soft (tangent hyperbolic) one, and investigate whether the same interesting physics -- particularly, relativistic backward scattering -- still occurs in such a structure for the case of sub-period transitions.
 
\section{Soft Potential Structure}
\begin{figure}[h!]
\begin{center}
\noindent
  \includegraphics[width=0.5\textwidth]{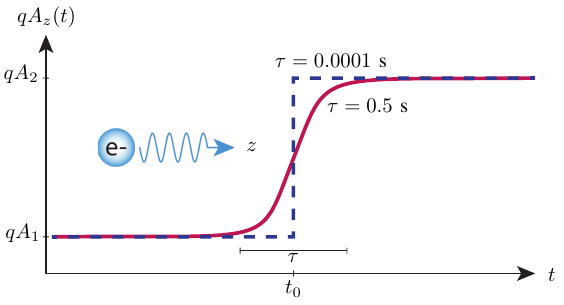}\vspace{-3mm}
  \caption{Smooth temporal potential step. The potential $A$ is directed along the propagation direction of the electron, $z$.}\label{fig:smooth_pot_str}
\end{center}
\end{figure}

\sss
\pati{Soft Potential Equation and Shape}
We consider the soft temporal potential step
\begin{equation}\label{eq:A_pot}
\begin{split}
    A_z(t)
    & = A_1 + \frac{A_2-A_1}{2} \left(1 + \tanh{\frac{t-t_0}{\tau}}\right) \\
    & = \frac{A_1+A_2\mathrm{e}^{2\frac{t-t_0}{\tau}}}{1+\mathrm{e}^{2\frac{t-t_0}{\tau}}}
\end{split}
\end{equation}
which is plotted in Fig.~\ref{fig:smooth_pot_str}. This potential varies between $A_1$ and $A_2$ at the transition time $t_0$ in time that is proportional to the transition time parameter $\tau$. 

\section{Dirac and Spinor's Component Equations}\sss
\pati{Dirac Equation}
According to~\cite{Ok_arXiv_2023}, the back-scattering occurring in the temporal potential step is a relativistic phenomenon. Therefore, the problem cannot generally be handled by the Schr\"{o}dinger equation; we must resort to the Dirac equation,
\begin{equation} \label{eq:Dirac}
    \left[\gamma^\mu(i\partial_\mu-qA_\mu)-m\right]\psi
    =0,
\end{equation}
where $\partial_{\mu}\equiv\left(\partial_{0},\partial_{i}\right)$ is the four-gradient, with $i=1,2,3$, and $\gamma^0=[0,I;I,0]$ and $\gamma^i=[0,-\sigma^i;\sigma^i,0]$ are the ($4\times{4}$) gamma matrices in the Weyl (Chiral) representation, with $\sigma^i$ being the ($2\times{2}$) Pauli matrices.

\pati{Spinor Wavefunction Ansatz}
We start with the spinor wavefunction ansatz
\begin{equation}\label{eq:psi}
    \psi=
    \begin{pmatrix}
        \varphi \\
        \vartheta
    \end{pmatrix}
    e^{i p z},
\end{equation}
where $p$ is the momentum of the electron. Inserting this ansatz and~\eqref{eq:A_pot} into~\eqref{eq:Dirac}, and solving the resulting system for $\varphi(t)$, yields a second-order differential equation in $\varphi(t)$. Subsequently, performing in that equation the change of variable $\zeta_\mathrm{e,l} = -\mathrm{e}^{ \pm 2\frac{t-t_0}{\tau}}$, where `$\mathrm{e}$' and `$\mathrm{l}$' refer to `earlier' and `later' than $t_0$, yields, after some algebraic manipulations, a new differential equation, this time in $\varphi_\mathrm{e,l}(\zeta_\mathrm{e,l})$ (one for $\mathrm{e}$ and one for $\mathrm{l}$). Introducing then the transformation $\varphi_\mathrm{e,l}(\zeta_\mathrm{e,l})=\zeta_\mathrm{e,l}^\mu(1-\zeta_\mathrm{e,l})^\nu f(\zeta_\mathrm{e,l})$ to eliminate the $\zeta_\mathrm{e,l}$ singularities at 0 and 1, finally yields the hypergeometric differential equation
\pati{Hypergeometric Differential Equation}
\begin{equation}
    \zeta_\mathrm{e,l} (1-\zeta_\mathrm{e,l}) \dv[2]{f}{\zeta_\mathrm{e,l}} + \left[c - (a+b + 1)\zeta_\mathrm{e,l} \right] \dv{f}{\zeta_\mathrm{e,l}} -\left (ab \right) f = 0,
\end{equation}
whose $f(\zeta_\mathrm{e,l})$ solution is a combination of Gauss' hypergeometric functions, $_2F_1(a,b;c;\zeta_\mathrm{e,l})$~\cite{gradshteyn2014table}, leading to the $\varphi_\mathrm{e,l}(\zeta_\mathrm{e,l})$ solution
\begin{subequations}\label{eq:phithet}
\begin{equation}
    \begin{split}
        \varphi_\mathrm{e,l}(\zeta_\mathrm{e,l}) 
            & = C_1{}_\mathrm{e,l} \zeta_\mathrm{e,l}^{\mu}(1-\zeta_\mathrm{e,l})^{\nu} {_2F_1}(a, b; c; \zeta_\mathrm{e,l}) \\ 
        & \quad+ C_2{}_\mathrm{e,l} \zeta_\mathrm{e,l}^{-\mu}(1-\zeta_\mathrm{e,l})^{\rho} {_2F_1}(a^{\prime}, b^{\prime}; c^{\prime}; \zeta_\mathrm{e,l}),
    \end{split}
\end{equation}
from which we also obtain
\begin{equation}
    \vartheta_\mathrm{e,l}=\frac{1}{m}\left[i\dv{\varphi_\mathrm{e,l}}{t}-(p-qA_z(t))\varphi_\mathrm{e,l}\right],
\end{equation}
\end{subequations}
from the Dirac equation [Eq.~\eqref{eq:Dirac}].

\pati{Boundary Condition}
Subjecting now~\eqref{eq:psi} with~\eqref{eq:phithet} to the boundary condition
\begin{equation}
    \psi_\mathrm{e}^D(t=t_0) = \psi_\mathrm{l}^D(t=t_0),
\end{equation}
where we have assumed reconversion from Weyl to Dirac spinors, determines the coefficients $C_1{}_\mathrm{e,l}$ and $C_2{}_\mathrm{e,l}$. We finally transform these functions into asymptotic forms ($t\rightarrow\pm\infty$) to obtain their plane-wave behavior, which leads to the solutions

\pati{Asymptotic Solutions}
\begin{subequations}
    \begin{equation}
    \begin{split}
        \psi_e^D(t\xrightarrow{}-\infty,z)
        & = \begin{pmatrix}
            \varphi_e^D \\
            \vartheta_e^D
        \end{pmatrix}
            e^{i p z} \\
            & = G_\mathrm{i} 
        \begin{pmatrix}
            1 \\
            \frac{E_1-m}{p-qA_1}
        \end{pmatrix}                          
        \mathrm{e}^{-i E_1(t-t_0)} e^{i p z},
    \end{split}
    \end{equation}
    \begin{equation}
        \begin{split}
            \psi_l^D(t\xrightarrow{}+\infty,z) &=
            \begin{pmatrix}
                \varphi_l^D \\
                \vartheta_l^D
            \end{pmatrix}
            e^{i p z} \\
            & = G_\mathrm{f} \begin{pmatrix}
                1 \\
                \frac{E_2-m}{p-qA_2}
            \end{pmatrix}
            \mathrm{e}^{-i E_2(t-t_0)} e^{i p z} \\
            & + G_\mathrm{b} \begin{pmatrix}
                1 \\
                \frac{-E_2-m}{p-qA_2}
            \end{pmatrix}
            \mathrm{e}^{i E_2(t-t_0)} e^{i p z},
            \end{split}
        \end{equation}
\end{subequations}
where
\begin{subequations}
    \begin{equation}
        G_\mathrm{i}=\frac{1}{\sqrt{2}}\frac{C_{2\mathrm{e}}}{m}\mathrm{e}^{\pi \frac{\tau}{2}E_1}\left(m + E_1 - (p-qA_1) \right),
    \end{equation}
    \begin{equation}
        G_\mathrm{f}=\frac{1}{\sqrt{2}}\frac{C_{1\mathrm{l}}}{m}\mathrm{e}^{-\pi \frac{\tau}{2}E_2}\left(m + E_2 - (p-qA_2) \right),
    \end{equation}
    \begin{equation}
        G_\mathrm{b}=\frac{1}{\sqrt{2}}\frac{C_{2\mathrm{l}}}{m}\mathrm{e}^{\pi \frac{\tau}{2}E_1}\left(m - E_2 - (p-qA_2) \right).
    \end{equation}
\end{subequations}
with
\begin{equation}    
    E_1 = \sqrt{(p-qA_1)^2 + m^2},
    \end{equation}
    and
    \begin{equation}
    E_2 = \sqrt{(p-qA_2)^2 + m^2}.
\end{equation}
\pati{Scattering probabilities and Amplitudes}
Finally, the later backward and forward scattering probabilities are obtained as
\begin{equation}
    F=\frac{f^2}{b^2+f^2}
    \quad\text{and}\quad
    B=\frac{b^2}{b^2+f^2},
\end{equation}
with the later forward and backward amplitudes
\begin{equation}
    f=\left|\frac{G_\mathrm{f}}{G_\mathrm{i}}\right|
    \quad\text{and}\quad
    b=\left|\frac{G_\mathrm{b }}{G_\mathrm{i}}\right| .
\end{equation}

\section{Results and Discussion}\sss

\begin{figure}[h]
\begin{center}
\noindent
  \includegraphics[width=0.5\textwidth]{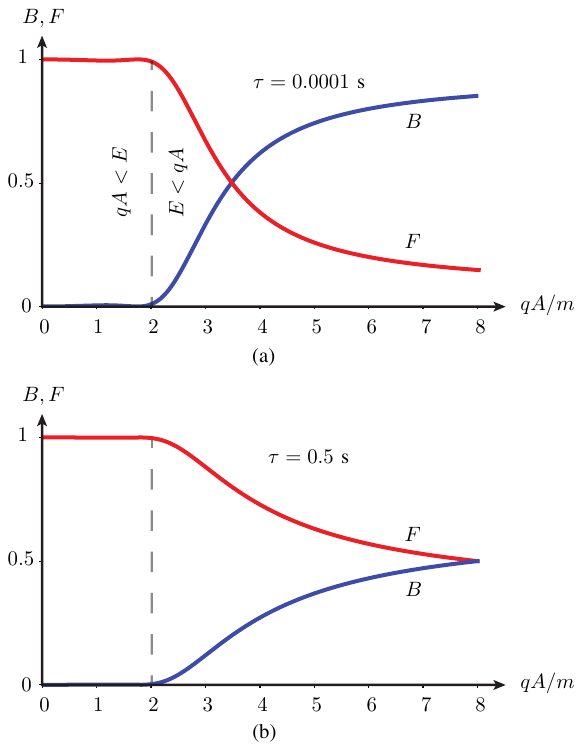}\vspace{-3mm}
  \caption{Scattering probabilities for the later backward ($B$) and forward ($F$) waves for (a)~a soft potential with $\tau=0.0001$~s in~\eqref{eq:A_pot}, which is equivalent to a sharp (Heaviside) potential, and (b)~a soft potential with $\tau=0.5$~s in~\eqref{eq:A_pot}. The (incident) energy to rest mass ratio $E/m=2$.}\label{fig:scatt_prob}
\end{center}
\end{figure}

\pati{Results and Discussion}
Figure~\ref{fig:scatt_prob} plots the scattering probabilities for the backward ($B$) and later forward ($F$) waves for two different transition widths in the soft potential in~\eqref{eq:A_pot} and Fig.~\ref{fig:smooth_pot_str}. Figure~\ref{fig:scatt_prob}(a) shows the case $\tau=0.0001$~s, which can in fact not be distinguished from the case of the sharp (acausal) step potential in~\cite{Ok_arXiv_2023}. Figure~\ref{fig:scatt_prob}(b) shows the case $\tau=0.5$~s. The fact that such a soft potential still supports a backward electron wave shows that this effect can easily be obtained in relatively slow and perfectly causal potentials.
 \vspace{-3mm}

\bibliographystyle{IEEEtran}
\bibliography{SMOOTH_STEP_Ok}

\end{document}